\documentclass[11pt,draftcls,onecolumn]{IEEEtran}

\makeatletter
\long\def\@makecaption#1#2{\ifx\@captype\@IEEEtablestring%
\footnotesize\begin{center}{\normalfont\footnotesize #1}\\
{\normalfont\footnotesize\scshape #2}\end{center}%
\@IEEEtablecaptionsepspace
\else
\@IEEEfigurecaptionsepspace
\setbox\@tempboxa\hbox{\normalfont\footnotesize {#1.}~~ #2}%
\ifdim \wd\@tempboxa >\hsize%
\setbox\@tempboxa\hbox{\normalfont\footnotesize {#1.}~~ }%
\parbox[t]{\hsize}{\normalfont\footnotesize \noindent\unhbox\@tempboxa#2}%
\else
\hbox to\hsize{\normalfont\footnotesize\hfil\box\@tempboxa\hfil}\fi\fi}
\makeatother

\usepackage[latin1]{inputenc}
\usepackage{times,amssymb,amsmath}
\usepackage{pstool}
\usepackage{subfigure}
\usepackage{multirow}
\usepackage{enumerate}
\usepackage{graphicx}
\usepackage{MnSymbol}
\usepackage{stfloats} 
\usepackage[table]{xcolor}
\usepackage[square,sort&compress,comma,numbers]{natbib}

\graphicspath{ {Figures/} }

\begin{document}

\title{A scalable architecture for distributed receive beamforming: analysis and experimental demonstration}

\author{
\IEEEauthorblockN{F.\ Quitin, A.T.\ Irish and U.\ Madhow}
}


\maketitle

\long\def\symbolfootnote[#1]#2{\begingroup%
\def\thefootnote{\fnsymbol{footnote}}\footnote[#1]{#2}\endgroup}
\symbolfootnote[0]{\hrulefill \\ 
F. Quitin is with the School of Electrical and Electronic Engineering, Nanyang Technological University (NTU), Singapore (fquitin@ntu.edu.sg). A.T. Irish and U. Madhow are with the Electrical and Computer Engineering Department, University of California, Santa Barbara (UCSB) (\{andrewirish, madhow\}@ece.ucsb.edu) \\
}

\begin{abstract}

We propose, analyze and demonstrate an architecture for scalable cooperative reception. In a cluster of $N+1$ receive nodes, one node is designated as the final receiver, and the $N$ other nodes act as amplify-and-forward relays which adapt their phases such that the relayed signals add up constructively at the designated receiver. This yields received SNR scaling linearly with $N$, while avoiding the linear increase in overhead incurred by a direct approach in which received signals are separately quantized and transmitted for centralized processing. By transforming the task of long-distance distributed receive beamforming into one of local distributed transmit beamforming, we can leverage a scalable one-bit feedback algorithm for phase synchronization. We show that time division between the long-distance and local links eliminates the need for explicit frequency synchronization. We provide an analytical framework, whose results closely match Monte Carlo simulations, to evaluate the impact of phase noise due to relaying delay on the performance of the one-bit feedback algorithm.  Experimental results from our prototype implementation on software-defined radios demonstrate the expected gains in received signal strength despite significant oscillator drift, and are consistent with results from our analytical framework.

\end{abstract}
\begin{keywords}
distributed MIMO, beamforming, cooperative reception, synchronization
\end{keywords}  

\section{Introduction}
\label{sec:intro}

Distributed MIMO (D-MIMO) refers to a broad class of techniques in which a group of cooperating nodes emulate a virtual antenna array, in order to
obtain performance gains similar to those provided by conventional centralized MIMO.  While it is difficult to scale centralized arrays to a
large number of elements due to size and weight considerations (especially at lower carrier frequencies), in principle, D-MIMO allows us to synthesize very large apertures using the natural geographic distribution of the cooperating nodes, and offer an approach to massive MIMO that sidesteps form factor constraints.  The opportunistic formation of D-MIMO clusters can also have
significant benefits in enhancing range/rate tradeoffs, especially in emergency and disaster relief scenarios. However, key difficulties in
translating D-MIMO from concept to practice are that the cooperating nodes have independent oscillators, each with stochastic drift, and that we cannot
rely on a regular array geometry in our signal processing algorithms.  Another important consideration in D-MIMO system design is that we would like our
architectures and algorithms to scale gracefully as the number of cooperating nodes increases, in order to approach the vision of arbitrarily large
virtual arrays.  In this paper, we address these issues in the context of distributed receive (D-Rx) beamforming.

\begin{figure}[ht]
	\centering
	\subfigure[Traditional Rx beamforming]{
	\includegraphics[width=8.5cm]{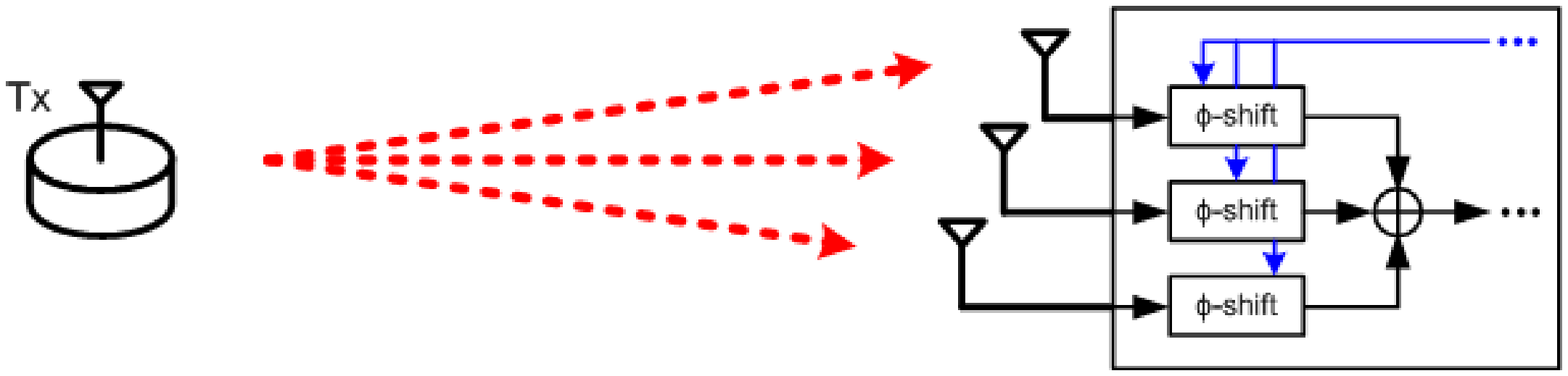}
	\label{subfig:architecture_1_traditional}
	}
	\subfigure[Naive D-Rx beamforming]{
	\includegraphics[width=8.5cm]{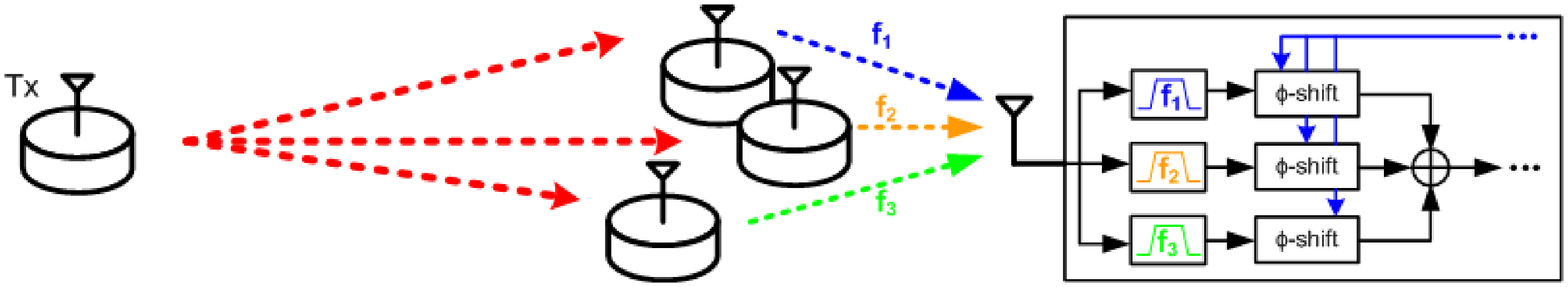}
	\label{subfig:architecture_2_naive}
	}
	\subfigure[Scalable D-Rx beamforming]{
	\includegraphics[width=8.5cm]{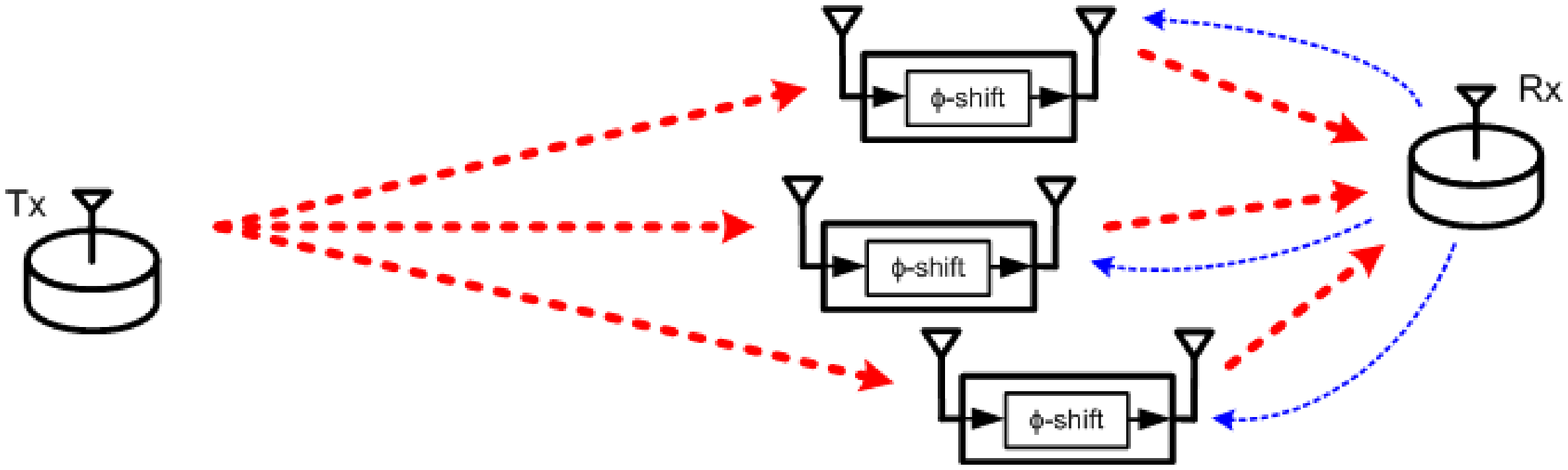}
	\label{subfig:architecture_3_scalable}
	}
	\caption{Rx beamforming with traditional MIMO, naive D-MIMO and scalable D-MIMO}
\end{figure}

In a D-Rx beamforming system, a cluster of nodes coherently combine their received signals in order to enhance the received signal-to-noise ratio (SNR).  In a centralized receive array, depicted in Figure~\ref{subfig:architecture_1_traditional} this is accomplished by routing signals from different receive antennas along wires, with phase shifts for coherent combining applied at RF or IF, or digitally at baseband, after downconversion and analog-to-digital conversion. An analogous approach for distributed receive beamforming,
shown in Figure~\ref{subfig:architecture_2_naive}, is for each node to send its received signal to a centralized processor (typically via a fast local wireless link), which then applies the appropriate phase shifts to achieve receive beamforming. With this approach, the cooperating nodes do not even have to be synchronized {\it a priori.}  The centralized processor has access to the received signal for each node, and hence can estimate relative frequency and phase offsets and then compensate for them when combining
the signals.  However, this direct approach does not scale to a large number of cooperating nodes, since the amount of local communication is proportional to the number of nodes. We therefore propose and investigate in this paper an alternative approach which attains scalability by using ``over the air'' coherent combining.

The proposed architecture (discussed in more detail in the next section) is depicted in Figure \ref{subfig:architecture_3_scalable}. A receive cluster of $N+1$ nodes wishes to enhance the SNR of a signal arriving
from a distant source. One of the receive nodes is designated as the final receiver, and the remaining
$N$ nodes relay their received signals, adjusting their phases such that the relayed signals combine coherently at the designated receiver.
This converts the task of D-Rx beamforming on the ``long link'' from source to receive cluster into one of distributed transmit (D-Tx) beamforming 
on the ``short link'' between the relays and the receiver.  This allows us to leverage prior work on scalable D-Tx beamforming,
while adapting to features and impairments peculiar to our relay-based architecture, in order to attain a scalable D-Rx system. 

The key features and contributions of this paper are summarized as follows:\\
\begin{enumerate}
	\item {\bf Scalable architecture: }By using amplify-forward relays in the receive cluster, we ensure that local communication overhead does not blow up with the number of cooperating nodes. We use a provably convergent one-bit feedback algorithm for distributed
	transmit beamforming in order to ensure that the relayed signals add up coherently at the receiver.
	\item {\bf Implicit frequency synchronization: }For time division between the long link and the short link as considered here,
	the frequency offsets of the relays ``cancel out'' on the long and short links, hence there is no need to synchronize the relays in
	frequency.
	\item {\bf Analytical performance characterization: }While frequency synchronization is not required, the delay between message
	reception on the long link and message relay on the short link leads to phase errors accumulating because of frequency and phase
	drift. We characterize the statistics of such phase errors as a function of local oscillator (LO) parameters, and then provide
	an analytical framework for determining their effect on the one-bit feedback algorithm.	Our analysis matches closely with Monte Carlo simulations, and shows that when the phase error gets large, only a fraction of the expected beamforming gain is achieved.
		\item {\bf Proof of concept: }The proposed architecture is implemented on a software-defined radio testbed, showing that the expected gains can be achieved with up to four relay nodes. By relaxing the system parameters, we are able to observe the performance of the one-bit feedback algorithm under larger phase errors experimentally, thereby verifying the insights from our analytical framework. 
\end{enumerate}

{\bf Related work: } Many information-theoretic analyses, ranging from three decades back \cite{Cover:TIT:1979, ElGamal:PROCIEEE:1980} to the present \cite{Borade:TIT:2007,Jin:TIT:2010,Gharan:TIT:2011}, rely on the concept of cooperative wireless communication, without explicitly addressing the fundamental bottleneck of synchronization between cooperating nodes. However, there has been significant recent progress on the problem of distributed synchronization, most of it in the context of distributed transmit beamforming.
Closed-loop synchronization techniques include explicit feedback \cite{Tu:VTC:2002,Ding:TWC:2008}, one-bit aggregate feedback \cite{Mudumbai:ISIT:2005,Mudumbai:ALL:2006,Mudumbai:TIT:2010}, implicit feedback using reciprocity \cite{Mudumbai:TWC:2007}, round-trip synchronization \cite{Ozil:ASIL:2007,Brown:TSP:2008} or two-way synchronization \cite{Preuss:CISS:2010,Preuss:TSP:2011}. Each of these synchronization techniques offers different trade-offs between complexity, coordination overhead and scalability to larger networks (see the review paper in \cite{Mudumbai:COMMAG:2009} and the discussion in \cite{Mudumbai:ASIL:2011}). Distributed MIMO techniques have also been investigated in the context of ``coordinated multipoint'' (CoMP) capabilities for 4G-LTE systems, where multiple base stations act as a distributed antenna array \cite{Irmer:COMMAG:2011}. Recent work \cite{Brown:TWC:2014} has shown that the overhead for D-Rx beamforming using the architecture in Figure~\ref{subfig:architecture_2_naive} can be reduced by heavily quantizing the information exchanged.  However, the local communication overhead still scales up with the
number of cooperating nodes, unlike our proposed approach.

There is also by now a significant body of research in analysis and simulation of amplify-and-forward relaying:
in \cite{Pun:TWC:2009}, the receiver broadcasts a single bit of information to each relay indicating whether it should participate in the communication, thereby selecting a set of relays who happen to be combining quasi-coherently; \cite{Jing:TIT:2009} considers network beamforming where each node has perfect channel state information; \cite{Zarifi:TSP:2011} considers network cooperation where each node computes beamforming weights based on local channel information only; and \cite{Zheng:SPL:2009,Zheng:TSP:2009} proposes a robust collaborative beamforming scheme based on partial channel state information. In \cite{Brown:TWC:2014}, nodes forward their message over a local area network, and it is shown that by selecting a subset of the receiving nodes, the performances are close to the ones obtained with optimal receive beamforming. 

While beamforming is implicit in the concept of ideal amplify-forward relay, other than our own prior conference paper  \cite{Quitin:VTC:2013}, to the best of our knowledge, the present paper is the first to identify the critical importance of amplify-forward relay in providing a scalable architecture for distributed receive beamforming, to model in detail the synchronization issues in implementing this approach, and to provide a proof of concept via our testbed.  A significant contribution of this paper beyond \cite{Quitin:VTC:2013} is that we provide
a detailed analytical framework for the impact of phase errors on the one-bit feedback algorithm.  We also extend the prototype in \cite{Quitin:VTC:2013} to a larger receive cluster, and provide a more detailed characterization of the effects of the system parameters on
experimental performance.

Distributed MIMO techniques have also been demonstrated with a wide variety of experimental prototypes. In \cite{Mudumbai:ALL:2006,Munkyo:MTT:2008,Sigg:ISSNIP:2009}, distributed transmit beamforming prototypes using wired feedback channels were presented for both RF and millimeter wave frequencies. A first fully wireless setup was presented in \cite{Rahman:IPSN:2012}, but this setup still used analog signaling for the feedback channel. A D-Tx beamforming prototype using full digital signaling and an extended Kalman filter for frequency synchronization was presented in \cite{Quitin:WoWMoM:2012, Quitin:GCOM:2012, Quitin:TWC:2013,Bidigare:ASIL:2012}. 
While the preceding papers focus on distributed transmit beamforming, multiuser distributed MIMO 
has been demonstrated in \cite{Balan:MOBICOM:2012}. However, the latter uses dedicated wired backhaul links to distribute information and feedback throughout the network. Similarly, CoMP experiments for 4G cellular systems have to date relied on 
dedicated backhaul links with low latency and high throughput, as well as assuming uninterrupted GPS connections to synchronize the various base stations. To the best of our knowledge, the present paper is the first work to analyze and prototype all-wireless distributed receive beamforming.

{\bf Outline of the paper: }Section \ref{sec:architecture} discusses our system model, the challenges of achieving synchronization at multiple levels in such a system, and points
out the implicit frequency synchronization achieved by the design considered here. In Section \ref{sec:phase_error}, we characterize the phase error accumulating due to relaying delay. Section \ref{sec:one_bit_fb} presents the phase synchronization algorithm and studies the performance of the this algorithm under the presence of phase error. Finally, Section \ref{sec:impl_results} presents the implementation of our architecture on an experimental testbed, and shows some of the results obtained with our prototype. The prototype is run with different parameters to confirm the theoretical insights of the previous sections. Section \ref{sec:conclusion}
contains our conclusions. 

\section{System Model}
\label{sec:architecture}

While there are many possible design choices for the scalable D-Rx architecture depicted in Figure~\ref{subfig:architecture_3_scalable},
the specific choices in our modeling, analysis and prototyping are as follows.  We focus on narrowband signaling, with channels modeled
as complex gains. Each relay node receives the signal from the distant transmitter over the long link, applies a phase shift to the received signal, and forwards it to a central receiver over a short link. All relays forward the message to the receiver over the same frequency band, with over-the-air
combining at the receiver. We assume that the same frequency band is used, via time division, for the long link and the short link, which enables
implicit frequency synchronization.  The relays adapt their phases using the one-bit feedback algorithm \cite{Mudumbai:TIT:2010}, which also formed the basis for our prior prototyping of
distributed transmit beamforming in \cite{Quitin:TWC:2013}.  Note that the relays can be implemented in RF or in baseband, assuming sufficient ADC resolution.

The key challenge in coherent combining at the receiver is that signals emitted from relays with independent clocks and oscillators must line up.
Three levels of synchronization are required: frequency, phase and timing.  In the following, we describe our approach to each of these problems.

\subsection{Implicit frequency synchronization}
\label{subsec:implicit_freq_sync}

In D-MIMO systems, each terminal derives its RF signal from its own local oscillator~(LO) which carries a small but non-zero frequency offset with respect to those of the other nodes. This results in LO frequency offsets that can range from a few Hz to tens of kHz for poor quality oscillators. If the signals forwarded by the relay nodes to the central receiver have significant frequency offsets, the total signal at the receiver will exhibit constructive and destructive interference patterns. In order to avoid such behavior, it is important that the signals arriving from the different relay nodes have no frequency offset with respect to one another. Fortunately, we get this for free in our architecture, as discussed in the following.

While we have transformed our D-Rx problem into one of distributed transmit beamforming along the short link,
a key difference from pure transmit beamforming context as in \cite{Quitin:TWC:2013} is that we obtain \textit{implicit} frequency synchronization by virtue of 
time division between the long and short links. To see this, 
suppose that the LO frequency offset between the transmitter and final receiver is denoted by $f_0$, and that the offset between the transmitter and the $i$-th relay is given by $f_{i1}$. The LO frequency offset between the $i$-th relay and the final receiver is then given by $f_{i2}=f_0-f_{i1}$. Because each relay node forwards the signal from the receiver \textit{without decoding}, the forwarded signal still contains the LO frequency offset of the first link. It then follows that all messages arriving at the receiver will carry the same total LO frequency offset of $f_{i1}+f_{i2}=f_0$, which is independent of the relay node $i$. 
Of course, in practice, LO phase and frequency drift causes an accumulation of phase error over the time between reception on the long link and relay on the short link.
For the low-quality oscillators in our software-defined radio testbed, such effects are significant.  In the next two sections, therefore, we model such phase errors and analyze
their impact on performance.

\subsection{Phase synchronization with the one-bit feedback algorithm} \label{subset:1bit}

Phase synchronization is achieved by using the well-known one-bit feedback algorithm \cite{Mudumbai:TIT:2010}, which is an iterative, scalable technique that has been shown to converge to an optimum both theoretically and experimentally. The algorithm is a stochastic hill climbing procedure, with one iteration per cycle. Each relay adds a random phase perturbation to its current phase, and forwards its received message. The receiver monitors the received signal strength (RSS) of the received, over-the-air combined message, and broadcasts back a single bit to the relays indicating whether or not the RSS has increased (compared to the earlier maximum RSS). If the RSS has increased, each relay keeps the last random phase perturbation and the new maximum RSS becomes the current RSS; if not, each relay discards its previous random phase perturbation and the maximum RSS remains unchanged. In the next cycle, the whole process is repeated. It has been proven theoretically \cite{Mudumbai:TIT:2010} and with numerous prototypes \cite{Mudumbai:ALL:2006,Rahman:IPSN:2012,Quitin:TWC:2013} that with this algorithm, the RSS quickly converges to its maximum value. 

\medskip

This ideal version of the one-bit feedback algorithm does however suffer from LO phase drift, which, unfortunately, occurs in any real world implementation. Consider the following scenario: at some point during the one-bit feedback algorithm, a combination of phases at the relay nodes is achieved that maximizes the RSS. Subsequent iterations of the algorithm will therefore not be able to improve on the previously obtained maximum. However, phase noise will cause the relays' phase offsets to drift away from their ideal values. Over time, the phase offsets will accumulate more random error, the RSS will vary randomly at lower values, and the one-bit feedback algorithm will fail entirely. 
To avoid this undesirable behavior, the following change is applied to the one-bit feedback algorithm. Instead of comparing the RSS in a given cycle to the maximum RSS achieved in a \textit{all} previous cycles, the current RSS is compared to the maximum RSS in the $K$ previous cycles (where $K$ is a finite number). The algorithm will then be able to recover from phase drifts, as the RSS at each cycle is only compared to the $K$ previous RSS values. If at some point a combination of phases is obtained with an RSS that cannot be outperformed, $K$ cycles later this maximum RSS will be removed from the past RSS memory, making the algorithm robust against phase drifts.

\subsection{Implicit timing synchronization}

The third type of synchronization is temporal: the packets forwarded by the relays must arrive simultaneously at the receiver to avoid the effects of inter-symbol interference (ISI). In this work, we assume a narrowband channel, so that the differences in propagation delays are significantly smaller than the inverse bandwidth and can be neglected. In that case, determining packet boundaries on the long link accurately and delaying the forwarded packet for a fixed amount of time is sufficient to obtain precise message alignment at the receiver. Thus, we use the timings of the messages received by the relays on the long link to provide implicit timing for the messages relayed on the short link.  For wideband dispersive channels, more sophisticated strategies are required to handle ISI (e.g., OFDM with frame synchronization across relays), but this is left as a topic for future work. 

\section{Modeling phase errors}
\label{sec:phase_error}

We now model the phase error accumulated due to relaying delay. 
Suppose the transmitter sends a message at time instant $t_1$. We know that the LO frequency offset satisfies the following condition: 
\begin{equation}
	\label{eq:impl_freq_synch_1}
	f_{i1}(t_1)+f_{i2}(t_1) = f_{j1}(t_1)+f_{j2}(t_1) = f_0(t_1) \ \ \ \ \ \ \ \forall i,j
\end{equation}
However, in reality there is some time delay $T_{d}$ before the relays forward the message to the final receiver. The LO frequency offsets of the different relay nodes will drift independently, such that \eqref{eq:impl_freq_synch_1} is no longer satisfied, which can be expressed by
\begin{equation*}
	\label{eq:impl_freq_synch_2}
	f_{i1}(t_1)+f_{i2}(t_1+T_d) \neq f_{j1}(t_1)+f_{j2}(t_1+T_d) \ \ \ \ \ \ \ \forall i,j
\end{equation*}
As a result, the signals arriving at the receiver from different relays now have different LO frequency offsets, leading to fading in the total message. 

We now model the effect of LO drift in more detail. We show that it causes a zero mean Gaussian phase error, and compute the variance of the phase
error as a function of system parameters.

{\bf LO model: }We begin by reviewing the following state space LO model  \cite{Zucca:TUFFC:2005,Brown:ICASSP:2012}.
Suppose $\phi(t)$ and $\omega(t)$ are the LO phase and (angular) frequency offset at time $t$. The LO phase and frequency offset at time $t+T$ are given by: 
\begin{equation*}
	\label{eq:LO_model}
	\left[ \begin{array}{l} \phi(t+T) \\ \omega(t+T) \end{array} \right] = 
							\left[ \begin{array}{l l} 1 & T \\ 0 & 1 \end{array} \right] 
							\left[ \begin{array}{l} \phi(t) \\ \omega(t) \end{array} \right]+ \mathbf{n}(T)
\end{equation*}
where $\mathbf{n}(T)$ is the process noise vector, distributed as $\mathbf{n} \sim \mathcal{N}(0,\mathbf{Q}(T))$, which causes the LO phase and frequency offset to drift from their nominal values. The process noise covariance matrix $\mathbf{Q}(T)$ is modeled as \cite{Zucca:TUFFC:2005,Brown:ICASSP:2012}
\begin{equation*}
	\label{eq:LO_model_Q}
	\mathbf{Q}\left(T\right) = \omega_c^2 q_1^2 \left[ \begin{array}{c c} T & 0 \\ 0  & 0 \end{array} \right] 
				+ \omega_c^2 q_2^2 \left[ \begin{array}{c c} \frac{T^3}{3} & \frac{T^2}{2} \\	\frac{T^2}{2} & T \end{array} \right]
\end{equation*}
where $\omega_c$ is the carrier frequency and the parameters $q_1^2$ and $q_2^2$ are the process noise parameters that correspond to white frequency noise and random walk frequency noise, respectively. 

\medskip

\begin{figure}[ht]
	\centering
	\includegraphics[width=8.5cm]{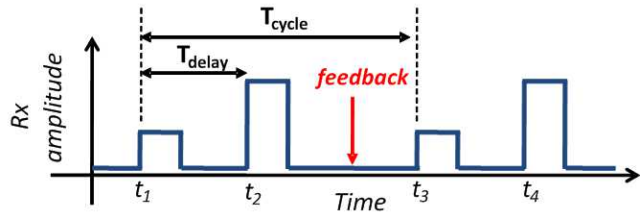}
	\caption{Received signal over two cycles.}
	\label{fig:time_rx_signal}
\end{figure}

{\bf Intra-cycle drift: } Consider the signal at the final receiver shown in Figure \ref{fig:time_rx_signal}. At time $t_1$, the distant transmitter sends its message to the relays (received with low amplitude), and at time $t_2$, the relays amplify and forward the message to the receiver (received with higher amplitude). The phase of the relayed signal at the receiver is then given by
\begin{align*}
	\label{eq:intra_cycle}
	\phi_i(t_2) &= \phi_{i1}(t_1) + \phi_{i2}(t_2) \\
	 						&= \phi_{i1}(t_1) + \phi_{i2}(t_1) + T_d \omega_{i2}(t_1) + n_{\phi_{i2}}(T_d) + \int\limits_{t_1}^{t_2} n_{\omega_{i2}}(\tau) d\tau
\end{align*}
where $\phi_{i1}$ and $\phi_{i2}$ denote the phase offset of the long and the short link, $\phi_{i}$ denotes the phase offset of the global link (from transmitter to receiver) through relay $i$, $n_{\phi_{i2}}$ is the LO phase noise over the short link and $n_{\omega_{i2}}$ is the LO frequency noise over the short link. The variance due to phase noise is then given by $\omega_c q_1^2 T_d$, whereas the variance due to the LO frequency noise is given by $\omega_c q_2^2 \frac{T_d}{3}$. The latter 
follows from the well-known result that for a standard Wiener process $W$, we have that $\int\limits_{a}^{b} W_t dt \sim \mathcal{N}(0,\frac{b-a}{3})$. The total phase error variance due to intra-cycle drift is thus modeled as a zero mean Gaussian variable with variance
\begin{equation}
	\label{eq:intra_cycle_var}
	\sigma_{\phi}^2 = \omega_c^2 q_1^2 T_d + \omega_c^2 q_2^2 \frac{T_d^3}{3}
\end{equation}
The increase in variance with relaying delay $T_d$ is intuitively reasonable: the longer the relays wait before forwarding the message, the larger the phase errors
accumulated due to the independent LO drifts at the different relays.

\medskip

{\bf Inter-cycle drift: } Figure \ref{fig:time_rx_signal} shows several relay cycles. At time $t_1$, the transmitter sends a first message, which is forwarded by the relays at time $t_2$. The receiver then computes its feedback message that is returned to the relay nodes; essentially, this message controls the phase $\gamma_i$ that is applied by each relay, such that the total phase $\phi_{i1}(t_1) + \phi_{i2}(t_2) + \gamma_i$ is equal for all links. Now consider a second cycle of the system: at time $t_3$, the transmitter sends another message, which is forwarded by the relays at time $t_4$. The total phase of the second message for link $i$ is given by
\begin{align}
	\phi_i(t_4) &= \phi_{i1}(t_3) + \phi_{i2}(t_4) + \gamma_i \nonumber \\
	 						&= \phi_{i1}(t_3) + \phi_{i2}(t_3) + T_d \omega_{i2}(t_3) + n_{\phi_{i2}}(T_d) + \int\limits_{t_3}^{t_4} n_{\omega_{i2}}(\tau) d\tau\ + \gamma_i \nonumber \\
	\label{eq:inter_cycle_3}
	 						&= \phi_{i1}(t_3) + \phi_{i2}(t_3) + T_d \left( \omega_{i2}(t_1) + n_{\omega_{i2}}(T_c) \right) + n_{\phi_{i2}}(T_d) + \int\limits_{t_3}^{t_4} n_{\omega_{i2}}(\tau) d\tau\ + \gamma_i
\end{align}
where $T_c$ is the cycle time. It can be seen that between $t_1$ and $t_3$ the LO frequency offset $\omega_{i2}$ has drifted, potentially causing the relay phase shift $\gamma_i$ to become outdated. Obviously, if the cycle period gets longer, this drift will be more severe, resulting in a larger phase error. From \eqref{eq:inter_cycle_3}, the total phase error due to both intra- and inter-cycle drift can be characterized by a zero-mean Gaussian process with variance 
\begin{equation}
	\label{eq:inter_cycle_var}
	\sigma_{\phi}^2 = \omega_c^2 q_1^2 T_d + \omega_c^2 q_2^2 \frac{T_d^3}{3} + \omega_c^2 q_2^2 T_d^2 T_c
\end{equation}
The first two terms of the equation are the same as for \eqref{eq:intra_cycle_var} and represent the intra-cycle drift, whereas the last term represent the inter-cycle drift. It can be concluded from \eqref{eq:inter_cycle_var} that, for fixed LO parameters $q_1^2$ and $q_2^2$, both the relay delay time $T_d$ and the cycle time $T_c$ should be kept low in order to maintain small phase errors. 

In the next section, we analyze the effect of phase errors on the performance of the one-bit feedback algorithm.
\section{One-bit feedback with phase errors}
\label{sec:one_bit_fb}

In Section \ref{sec:phase_error}, we showed that implicit frequency synchronization results in a zero-mean Gaussian phase error for each relay. Even for poor quality LOs, these phase errors are well below $2\pi$, and do not affect frequency synchronization. However, they do affect phase synchronization, and in this section, we develop an analytical framework to characterize its impact on the one-bit feedback algorithm
(with RSS memory of length $K$; see description in Section \ref{subset:1bit}).  In Section \ref{sec:impl_results}, the effects of increased phase errors is investigated experimentally. 

\medskip

The key steps in our derivation are the following: 
\begin{enumerate}
	\item We determine the probability that the \textit{noisy} phases cause an RSS increase, conditioned on the fact that the \textit{noiseless} phases cause an RSS increase, over a single cycle of the algorithm (Section~\ref{subsec:one_bit_fb_gauss_phase_noise}). 
	\item In the case of finite memory, we define $K$ states $S_k$ as being the situation where the maximum RSS was achieved $k$ cycles ago, and model the transitions between states as a Markov chain. Following arguments similar to the ones in Section~\ref{subsec:one_bit_fb_gauss_phase_noise}, we compute the phase transition probabilities and derive the probability of being in each state $S_k$ (Section~\ref{subsec:one_bit_fb_mem_K_gauss_ph_noise}). 
	\item We compute the RSS drift -- the expected deviation in RSS at each iteration in the algorithm -- when operating with noisy phases, both for positive and negative feedback, conditioned on the current state (Section~\ref{subsec:one_bit_fb_mem_K_gauss_ph_noise}). 
	\item Using the probabilities of being in each state $S_k$ and the RSS drift conditioned on state $S_k$, we determine the total RSS drift when running the one-bit feedback algorithm with noisy phases (Section~\ref{subsec:one_bit_fb_mem_K_gauss_ph_noise}). 
\end{enumerate}

\subsection{One-bit feedback algorithm with Gaussian phase noise over a single cycle}
\label{subsec:one_bit_fb_gauss_phase_noise}

We start by investigating the effect of Gaussian phase noise over a single cycle. For $N$ nodes, we define the following normalized RSS values
\begin{align*}
 y 						&=\frac{1}{N}\sum\limits_{i=1}^{N}\left|e^{j(\phi_i)}\right| \\
 y_{\delta} 	&=\frac{1}{N}\sum\limits_{i=1}^{N}\left|e^{j(\phi_i+\delta_i)}\right| \\
 y_n 					&=\frac{1}{N}\sum\limits_{i=1}^{N}\left|e^{j(\phi_i+n_i)}\right| \\
 y_{\delta n} &=\frac{1}{N}\sum\limits_{i=1}^{N}\left|e^{j(\phi_i+\delta_i+n_i')}\right|
\end{align*}
that represent the noiseless RSS before random phase perturbation, the noiseless RSS after random phase perturbation, the noisy RSS before random phase perturbation and the noisy RSS after random phase perturbation, respectively. In these equations, $\delta_i$ is the random phase perturbation applied at each node, and $n_i$, $n_i'$ are the phase noises (before and after random phase perturbation) at node $i$. 

The problem at hand is then to determine when the one-bit feedback performs as well in the noisy case as in the noiseless case. To begin, we define
\begin{align*}
 U &\triangleq y_{\delta n}-y_{n}  \\
 V &\triangleq y_{\delta}-y
\end{align*}
which denote the RSS increment after a phase perturbation in the noisy and noiseless cases, respectively. In addition, let us define the state of the one bit feedback process as the value $y$ of the RSS in the noiseless case. Conditioned on the feedback process state, the probability of successful detection of a phase improvement/deterioration is then
\begin{align*}
 &\mathbb{P}\left[U>0|V>0,y\right]  \\
 &\mathbb{P}\left[U<0|V<0,y\right] 
\end{align*}
Clearly, the random variables $U$ and $V$ are not independent, so to determine $\mathbb{P}\left[U>0|V>0,y\right]$ and $\mathbb{P}\left[U<0|V<0,y\right]$, one must first determine the joint probability distribution of $(U,V)$. To proceed, we leverage the results derived in Conjecture 1 in \cite{Mudumbai:TIT:2010}, where statistical mechanics arguments were used to determine that that the distribution of the phases around their mean follow an Exp-Cosine distribution. Given this framework, we show in Appendix~\ref{sec:joint_prob_U_V} that the joint probability distribution of $(U,V)$ can be approximated by the following bivariate Gaussian distribution 
\begin{align}
\label{eq:joint_distr_UV}
\left[\begin{array}{c}
	U \\ V
\end{array} \right] \sim \mathcal{N}\left( 
\left[\begin{array}{c}
	(\chi_{\delta n}-\chi_n)y \\ (\chi_{\delta}-1)y 
\end{array} \right], 
\left[\begin{array}{c c}
	\frac{1-\chi_{\delta n}^2-\rho_{\delta n}\kappa(y)+1-\chi_{n}^2-\rho_{n}\kappa(y)}{2N} & \frac{\chi_n-\chi_{\delta n}\chi_{\delta} - \rho_{2\delta n}\kappa(y)}{2N} \\
	\frac{\chi_n-\chi_{\delta n}\chi_{\delta} - \rho_{2\delta n}\kappa(y)}{2N} & \frac{1-\chi_{\delta}^2-\rho_{\delta}\kappa(y)}{2N} 
\end{array} \right]
 \right)
\end{align}
where $\chi_{\delta}=\mathbb{E}[\cos(\delta_i)]$, $\chi_{n}=\mathbb{E}[\cos(n_i)]$ and $\chi_{\delta n}=\mathbb{E}[\cos(\delta_i+n_i')]$, where $\rho_{\delta}=\chi_{\delta}^2-\mathbb{E}[\cos(2\delta_i)]$, $\rho_{n}=\chi_{n}^2-\mathbb{E}[\cos(2n_i)]$ and $\rho_{\delta n}=\chi_{\delta n}^2-\mathbb{E}[\cos(2(\delta_i+n_i))]$, where $\rho_{2\delta n} = \chi_{\delta n}\chi_{\delta} - \mathbb{E}[\cos(2\delta_i+n_i)]$, and where the term $\kappa(y)=\frac{1}{N}\sum\limits_{i=1}^{N}\mathbb{E}[e^{2j\phi_i}]$ depends on $y$ only, and can be approximated by $\kappa(y)=e^{-4(1-y)}$ for large $y$. From \eqref{eq:joint_distr_UV}, one can easily compute the probability that the one-bit feedback algorithm is successfully able to detect a phase improvement $\mathbb{P}\left[y_{\delta n}>y_{n}|y_{\delta}>y,y\right]$ or a phase deterioration $\mathbb{P}\left[y_{\delta n}<y_{n}|y_{\delta}<y,y\right]$ in the noisy case. 

\medskip

\begin{figure}[ht]
	\centering
	\includegraphics[width=10cm]{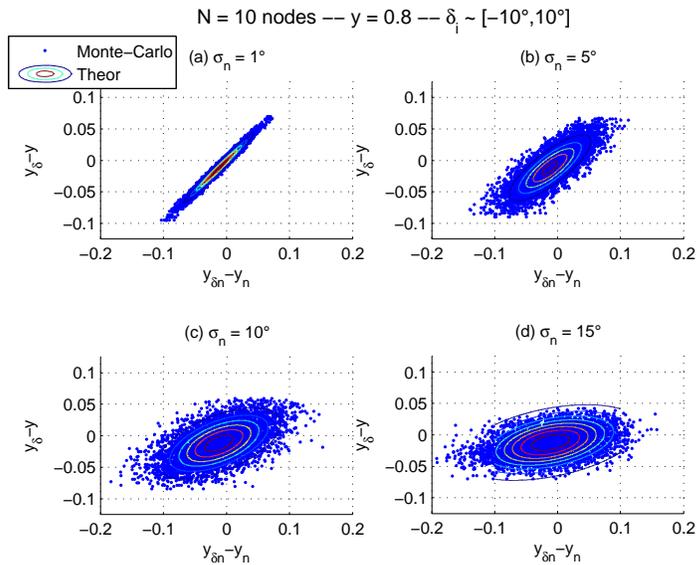}
	\caption{Comparison between Monte-Carlo simulations and theoretical model for various phase noise variances and $N=10$ nodes. The random phase perturbations are chosen uniformly from the discrete set $\{-10^{\circ},+10^{\circ}\}$. }
\label{fig:verify_distr_unif}
\end{figure}

The joint distribution \eqref{eq:joint_distr_UV} has been compared with Monte-Carlo simulations and is shown to match the simulations very well for as few as 10 nodes. In these simulations, the random phase perturbation was chosen uniformly from the discrete set $\{-10^{\circ},+10^{\circ}\}$, as to match our experiments in Section~\ref{sec:impl_results}. Note that other distributions for the random phase perturbations yield similar conclusions. The phase noise is drawn from a zero-mean Gaussian distribution with variance $\sigma_n^2$. It can be seen in Figure~\ref{fig:verify_distr_unif} that, for low phase noise, $U$ and $V$ are highly correlated: a successful decision in the noiseless case gives way to an identical decision in the noisy case. As the phase noise increases, $U$ and $V$ become less correlated, and the probability that the one-bit feedback algorithm makes a correct decision (with respect to the noiseless case) becomes lower. 

\subsection{One-bit feedback algorithm with Gaussian phase noise and an RSS memory of length $K$}
\label{subsec:one_bit_fb_mem_K_gauss_ph_noise}

\begin{figure}[ht]
	\centering
	\includegraphics[width=8.5cm]{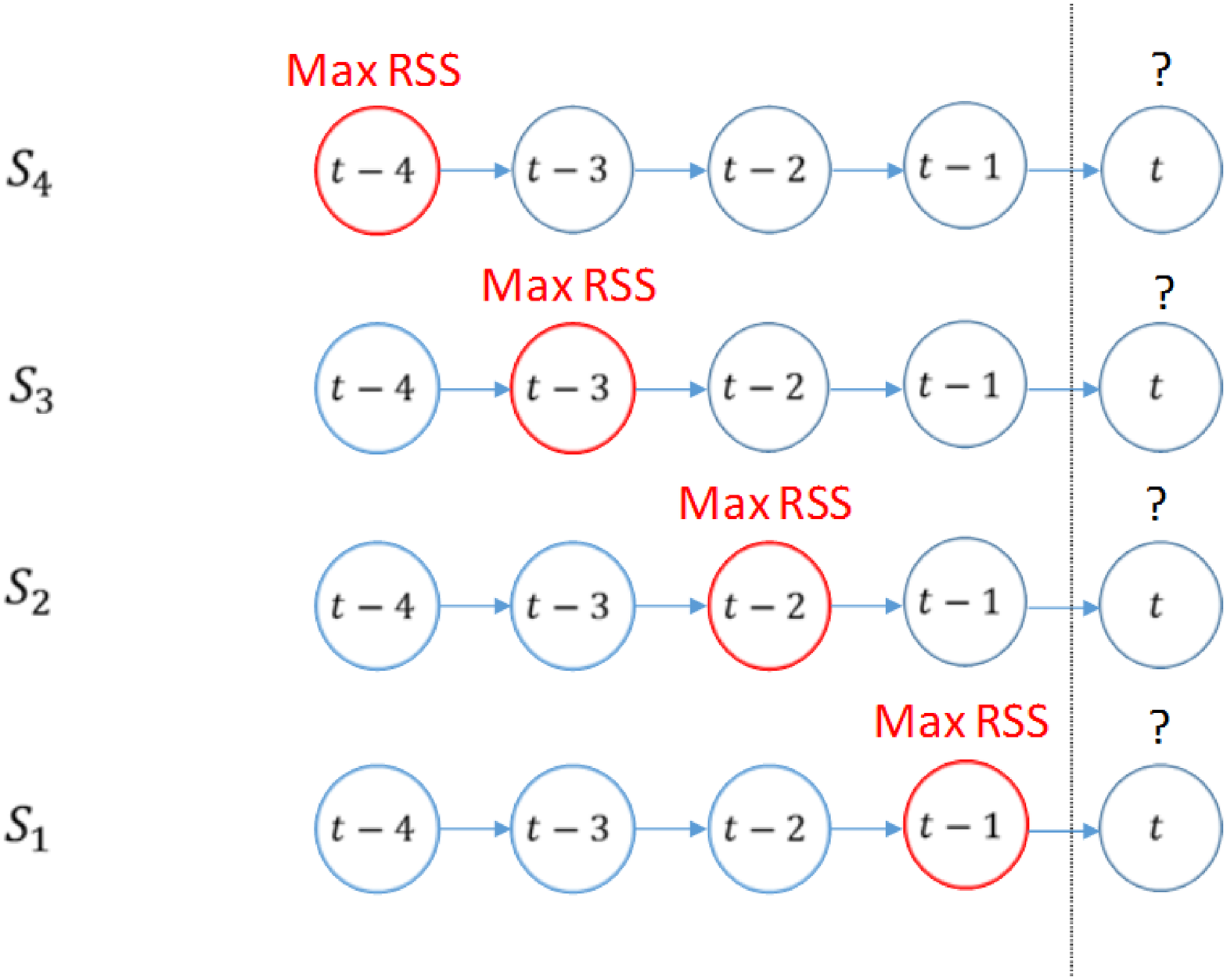}
	\caption{Different states of the system for a memory of size $K=4$.  }
	\label{fig:illustration_states}
\end{figure}

As mentioned, a practical implementation of the one-bit feedback algorithm requires the use an RSS memory of finite size. We therefore consider the case where the one-bit feedback algorithm compares the current RSS at time $t$ with the RSS of the past $K$ cycles. In the following framework, we define the state $S_k$ as the state where the maximum RSS during the $K$ previous cycles was obtained during cycle $t-k$, as shown in Figure~\ref{fig:illustration_states}. At each time instant, we are in one of the $K$ possible states $S_k$. 

We are interested in computing the total RSS drift, that is the average RSS increment at time $t$ conditioned on the current RSS level $y$. We make the following simplifying assumption: if we are in state $S_k$ at time $t$, the RSS drift is \textit{statistically independent} of the feedback before time $t-k$. In other words, the RSS drift is only dependent on the feedback obtained between time $t-k$ and time $t$. This can be verified as follows: imagine the following time instants $\tau_0< \tau_1< t-k <\tau_2\leq t$, where $\tau_0$ and $t-k$ correspond to two time instants when there was a phase update. Instant $\tau_0$ is the reference point at time $\tau_1$, and $t-k$ is the reference point at time $\tau_2$ and time $t$. Since the random phase perturbations and phase noises are independent across iterations, we can show that the covariance between $U_{\tau_1}$ and $U_{\tau_2}$ and the covariance between $V_t$ and $U_{\tau_1}$ is equal to (using the notations defined in Appendix~\ref{sec:joint_prob_U_V})
\begin{align*}
\text{Cov}\left[U_{\tau_1},U_{\tau_2}\right] &= \mathbb{E}\left[ \left(x_{\Re,\delta n,\tau_1}-x_{\Re, n,\tau_0}\right)\left(x_{\Re,\delta n,\tau_2}-x_{\Re,t-l}\right)^* \right] = 0\\
\text{Cov}\left[V_t,U_{\tau_2}\right] &= \mathbb{E}\left[ x_{\Re,\delta,t} \left(x_{\Re,\delta n,\tau_1}-x_{\Re,n,\tau_0}\right)^* \right] = 0\\
\end{align*} 
which shows that the RSS drift at time $t$ is independent of the feedback before $t-k$.

\medskip

Given the previous assumption, the (noiseless) RSS drift conditioned on the current state $S_k$ is given by 
\begin{align*}
\text{Drift}(RSS_t|S_k,y) &= \mathbb{E}\left[V_t|\text{feedback since }t-k,y\right] 
\end{align*}
The total RSS drift with a memory of length $K$ is then given by: 
\begin{align}
\text{Drift}(RSS_t|K,y) &= \sum\limits_{k=1}^{K}\mathbb{P}\left(S_k|y\right) \cdot \text{Drift}(RSS_t|S_k,y) \nonumber \\
 &= \sum\limits_{k=1}^{K} \mathbb{P}\left(S_k|y\right) \cdot \left[ \mathbb{P}(U_t>0|S_k)\text{Drift}(RSS_t|S_k,U_t>0,y) + \mathbb{P}(U_t<0|S_k)\text{Drift}(RSS_t|S_k,U_t<0,y) \right] \label{eq:rss_drift_1}
\end{align}

\medskip

\begin{figure}[ht]
	\centering
	\includegraphics[width=8.5cm]{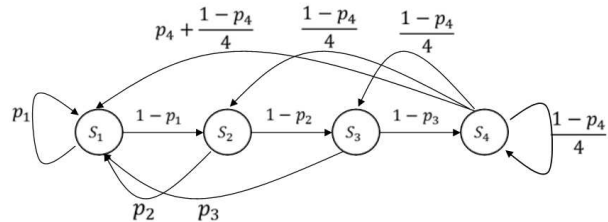}
	\caption{Markov chain modeling the state transitions for a memory of size $K=4$.  }
	\label{fig:markov_chain}
\end{figure}

We begin by determining the probabilities $\mathbb{P}\left(S_k|y\right)$ of being in a given state $k$. We model the state transitions with the Markov chain shown in Figure~\ref{fig:markov_chain}. 
In each state, a positive feedback will bring the algorithm back in state $S_1$, whereas a negative feedback will cause the algorithm to transition from state $S_k$ to state $S_{k+1}$. For the final state $S_K$, a negative feedback will cause to algorithm to transition to any state with uniform probability, as the noisy RSS between sates $t,t-1,...,t-K+1$ are i.i.d. The Markov chain state transition matrix is given by (for a memory of length 4)
\begin{align}
\mathbf{P} = \left[ 
\begin{array}{c c c c}
	p_1 &1-p_1 &0 &0 \\
	p_2 &0 &1-p_2 &0 \\
	p_3 &0 &0 &1-p_3 \\
	p_4+\frac{1-p_4}{4} &\frac{1-p_4}{4} &\frac{1-p_4}{4} &\frac{1-p_4}{4}
\end{array}
 \right] \label{eq:P_matrix}
\end{align}
By looking at the stationary distribution $\mathbf{\pi}$ of the Markov chain, we can determine the probability of being in each state. The stationary distribution $\mathbf{\pi}$ is the left eigenvector of $\mathbf{P}$ that correspond to the eigenvalue $\lambda=1$: 
\begin{align}
\mathbf{\pi}^T\mathbf{P} = \mathbf{\pi}^T \label{eq:Markov_stat}
\end{align}
which will depend on the state transition probabilities $p_k$ and the memory length $K$. 

\medskip

\begin{figure}[ht]
	\centering
	\includegraphics[width=8.5cm]{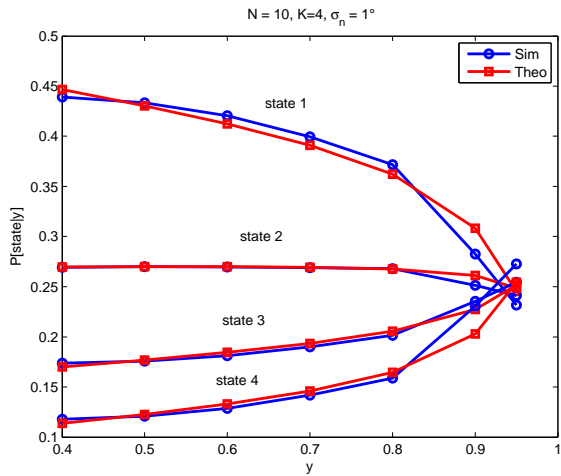}
	\caption{Probability of being in each state,  for $N=10$ nodes, a memory of size $K=4$, a random phase perturbation of $\delta\sim\mathcal{U}[-10^{\circ},10^{\circ}]$, and a phase noise of $\sigma_n=1^{\circ}$.  }
	\label{fig:prob_state}
\end{figure}

We now determine the state transition probabilities $p_k$, which define the probability of a positive feedback at a given state $S_k$. The probability $p_k$ is defined as
\begin{align}
p_k &= \mathbb{P}\left[ U_t>0 | S_k; y\right] \nonumber \\
		&= \mathbb{P}\left[ U_t>0 | U_{t-1}<0,...,U_{t-k+1}<0; y\right] \label{eq:p_k}
\end{align}
Using similar arguments as in Section~\ref{subsec:one_bit_fb_gauss_phase_noise}, we model the random variable $\left[U_t,U_{t-1},...,U_{t-k+1}\right]^T$ as a multivariate Gaussian random variable. The mean and variances of each term have already been computed in Section \ref{subsec:one_bit_fb_gauss_phase_noise}, and the covariance $\rho_{UU}$ between two variables $U_{t_1}$ and $U_{t_2}$ is given by (using arguments similar to the one in Appendix~\ref{sec:joint_prob_U_V})
\begin{align*}
\rho_{UU} &= \mathbb{E}\left[\left(x_{\Re,\delta n,t_1}-x_{\Re,\delta n,t-l}\right) \left(x_{\Re,\delta n,t_2}-x_{\Re,\delta n,t-l}\right)\right] \\
					&= \mathbb{E}\left[ x_{\Re,\delta n,t-l}^2 \right] \\
					&= \frac{1-\chi_{\delta n}^2 - \rho_{\delta n}\kappa(y)}{2N}
\end{align*}
Note that this term is identical for all possible values of $t_1$ and $t_2$. Once we know the mean and full covariance matrix of the multivariate Gaussian random variable $\left[U_t,U_{t-1},...,U_{t-k+1}\right]^T$, we can use determine the probability $p_k$ given in \eqref{eq:p_k} by computing the multivariate cumulative density function. By filling the values of $p_k$ in \eqref{eq:P_matrix} and \eqref{eq:Markov_stat}, one can compute the probability $\mathbb{P}(S_k)$ of being in state $k$. Figure~\ref{fig:prob_state} shows the probability of being in each state for $N=10$ nodes and a memory of size $K=4$, and compares our theoretical model with Monte-Carlo simulations. It can be seen that there is a good match between theory and simulations. 

\medskip

We now compute the RSS drift terms of equation \eqref{eq:rss_drift_1}. The drift term for positive feedback is defined as 
\begin{align}
	\text{Drift}(RSS_t|S_k,U_t>0;y) = \mathbb{E}\left[V_t|U_t>0,U_{t-1}<0,...,U_{t-k+1}<0;y \right] \label{eq:drift_term_1}
\end{align}
We again consider the Gaussian multivariate distribution $[V_t,U_t,U_{t-1},...,U_{t-k+1}]$. The mean and variances of all the terms have been computed previously, as have the covariance between $U_{t_1}$ and $U_{t_2}$ and the covariance $\rho_{UV}$ between $U_t$ and $V_t$. The covariance $\rho_{U_{\tau}V_t}$ between $V_t$ and $U_{\tau}$ is given by (for $t-l<\tau<t$)
\begin{align*}
\rho_{U_{\tau},V_t} &= \mathbb{E}\left[ \left(x_{\Re,\delta n,\tau}-x_{\Re,\delta n,t-l}\right) x_{\Re,\delta,t} \right] \\
					&= \mathbb{E}\left[\left(x_{\Re,\delta n,\tau}-x_{\Re,\delta n,t-l}\right)\right] \mathbb{E}\left[x_{\Re,\delta,t} \right] \\
					&= 0
\end{align*}
The covariance matrix of the multivariate Gaussian distribution is then given by
\begin{align*}
\mathbf{\Sigma} = \left[ 
\begin{array}{l l l l l}
	\sigma_V^2 &\rho_{UV} &0 &\cdots &0 \\
	\rho_{UV} &\sigma_U^2 &\rho_{UU} &\cdots &\rho_{UU} \\
	0 &\rho_{UU} &\sigma_U^2 &\cdots &\rho_{UU} \\
	\vdots  &\vdots  &\vdots  &\ddots &\vdots \\
	0 & \rho_{UU} &\rho_{UU} &\cdots &\sigma_U^2	
\end{array}
 \right] 
\end{align*}
Computing the RSS drift term in \eqref{eq:drift_term_1} then boils down to determining the mean of a truncated multivariate Gaussian distribution, which can be done efficiently with Monte-Carlo integral computation.

For the case of negative feedback, and for all states $k=1,...,K-1$, the RSS drift is equal to zero, as the one-bit feedback algorithm does not change the combination of phases: 
\begin{align}
	\text{Drift}(RSS_t|S_k,U_t<0;y) = \mathbb{E}\left[V_t|U_t<0,U_{t-1}<0,...,U_{t-k+1}<0;y \right] = 0 \label{eq:drift_term_3}
\end{align}
Note that this equality does not hold for $k=K$. If $U_t<0$, the previous maximum RSS then ``slips out'' of the past RSS memory, and in the next iteration a new maximum RSS will be considered. By definition, this will cause a change in true RSS drift, which can be expressed as
\begin{align}
	\text{Drift}(RSS_t|S_K,U_t<0;y) &= \mathbb{P}\left[\text{Max. noisy RSS at } t-\tau \right]   \mathbb{E}\left[V_{t-\tau}|U_{t-\tau}<0,\lbrace U_{t-l}<U_{t-\tau}\rbrace _{K-1\leq\tau\leq t,l\neq\tau};y \right] \nonumber \\
	&= \mathbb{E}\left[V_{t}|U_t<0,U_{t-1}<U_t,...,U_{t-K+1}<U_t;y \right] \label{eq:drift_term_4}
\end{align}
where this last equation is obtained by symmetry arguments. Equation \eqref{eq:drift_term_4} can be determined by yet another multivariate Gaussian distribution with variates $[V_t,U_t,U_{t-1},...,U_{t-K+1}]^T$. The elements of the mean and covariance matrix of this multivariate Gaussian distributions have all been computed previously, and the solution of \eqref{eq:drift_term_4} can be obtained through Monte-Carlo integral computation. 

\medskip

\begin{figure}[ht]
	\centering
	\includegraphics[width=8.5cm]{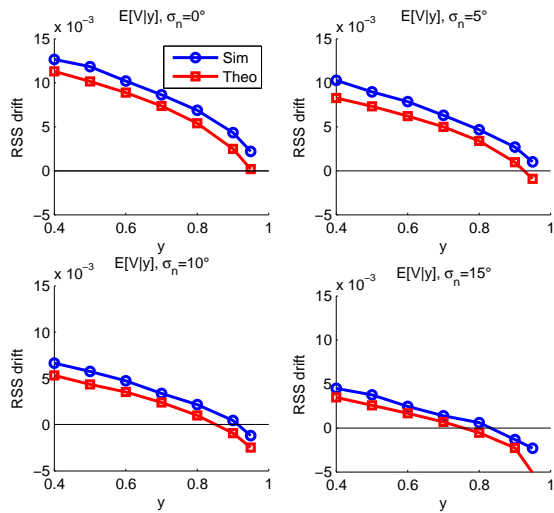}
	\caption{Theoretical and simulated RSS drift for $N=10$ nodes, a memory of size $K=4$ and a random phase perturbation of $\delta\sim\mathcal{U}[-10^{\circ},10^{\circ}]$, for various phase noise values.  }
	\label{fig:RSS_drift_july}
\end{figure}

By combining the different RSS drift terms \eqref{eq:drift_term_1}, \eqref{eq:drift_term_3} and \eqref{eq:drift_term_4} into equation \eqref{eq:rss_drift_1}, the total RSS drift can be computed. Figure~\ref{fig:RSS_drift_july} shows the RSS drift for various phase noise values, both with Monte-Carlo simulations and with our theoretical model. It can be seen that there is a good correspondence between the simulated and the theoretical curves. It should be noted here that to obtain stable Monte-Carlo curves, a large number of realizations must be generated, making the computational requirements quite expensive. In contrast, our theoretical analysis is fairly efficient and does not require significant computation power. 
From Figure~\ref{fig:RSS_drift_july} it can be seen that when the phase noise becomes larger than the random phase perturbation, the RSS drift eventually becomes negative. This means that in steady-state, the RSS will converge to a value of $y$ smaller than 1, and not achieve the maximum possible RSS. For a phase noise of $\sigma_n=15^{\circ}$, the RSS will only reach 80\% of the maximum achievable RSS. 

\begin{figure}[ht]
	\centering
	\includegraphics[width=8.5cm]{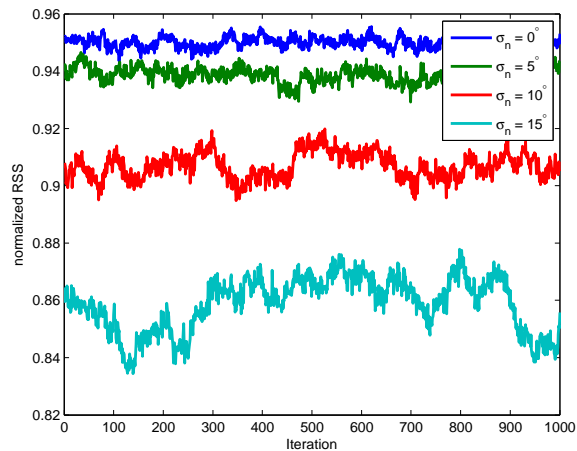}
	\caption{Normalized RSS when running the one-bit feedback algorithm, averaged over 100 runs. It can be seen that as the phase noise increases, the steady-state average RSS decreases. The average normalized RSS can be predicted by the zero-crossing of the RSS drift in Figure~\ref{fig:RSS_drift_july}. }
	\label{fig:avg_runs}
\end{figure}

Our analysis is confirmed by the simulation results in Figure~\ref{fig:avg_runs}. Here, the normalized RSS is plotted versus time when running the one-bit feedback algorithm. The normalized RSS has been averaged over 100 simulation runs, and the first 1000 iterations are not plotted in Figure~\ref{fig:avg_runs} (in order to focus only on the steady-state convergence values). It can be seen that the normalized RSS does converge at a value that is predicted by the zero-crossing of the RSS drift in Figure~\ref{fig:RSS_drift_july}. In Section~\ref{sec:impl_results} we will show that our experimental testbed starts failing when the phase noise gets larger than the random phase perturbations. 

\section{Experimental demonstration}
\label{sec:impl_results}

\subsection{Software-defined radio testbed}

The proposed architecture was implemented on a software-defined radio testbed using six USRP RF and baseband boards \cite{usrp:2013}. We use a mix of USRP-2 and USRP-N200 baseband boards, and WBX 50-2200~MHz RF daughterboards. Each USRP was connected to a host laptop that performed the computation using GNU~Radio software \cite{gnuradio:2013}. Our software is available for download online \cite{website_dTxBeamf}.  

\medskip

\begin{figure}[h]
	\centering
	\subfigure[Relay node]{
	\includegraphics[width=8.5cm]{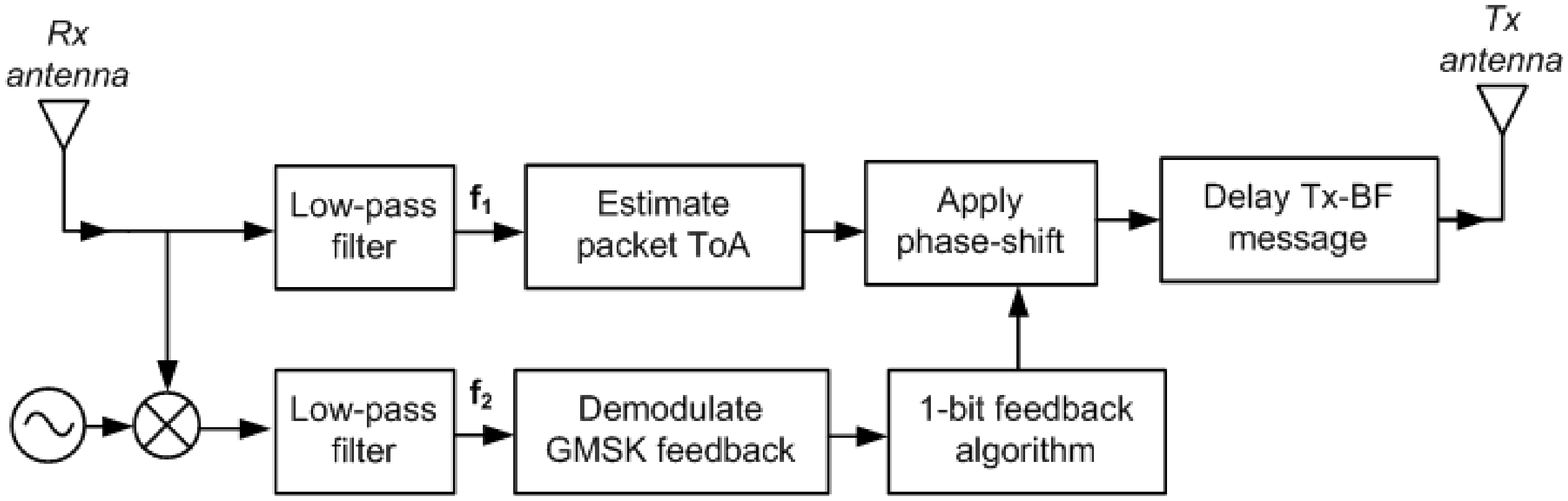}
	\label{subfig:block_diagram_relay}
	}
	\subfigure[Receive node]{
	\includegraphics[width=8.5cm]{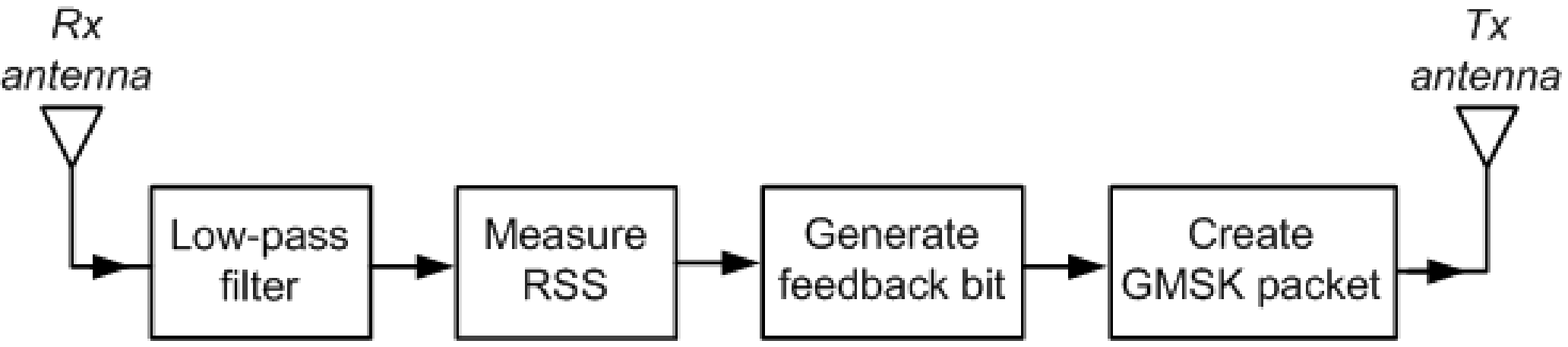}
	\label{subfig:block_diagram_rx}
	}
	\caption{Block-diagram for relay and receiver node. }
\end{figure}

One USRP was used as a transmit node (sending packets that contain only a pilot tone), one USRP was used as the final receiver, and up to four USRPs were used as relay nodes. A block diagram of the relay nodes is shown in Figure~\ref{subfig:block_diagram_relay}. The relay nodes receive the message from the transmitter, add a phase shift to the received message, and wait for a fixed amount of time $T_d$ before amplifying and forwarding the message to the final receiver (this delay $T_d$ needs to be identical for all relays for the messages to add up without ISI). The message is forwarded over the same frequency band as it is the one used by the transmitter, which was set to 908~MHz. Additionally, the relay nodes also listen to the feedback message over the feedback channel (at 928~MHz) to determine the phase shift to be applied using the one-bit feedback algorithm. The sample rate of all the nodes was set to 200~kHz. 
The receiver, shown in Figure~\ref{subfig:block_diagram_rx}, receives the combined messages of all relay nodes, computes its single bit of feedback and broadcasts this single bit back to all relays over the feedback channel. The single bit is embedded in a GMSK-modulated packet. 

\subsection{Experimental results}

\begin{figure}[ht]
	\centering
	\includegraphics[width=8.5cm]{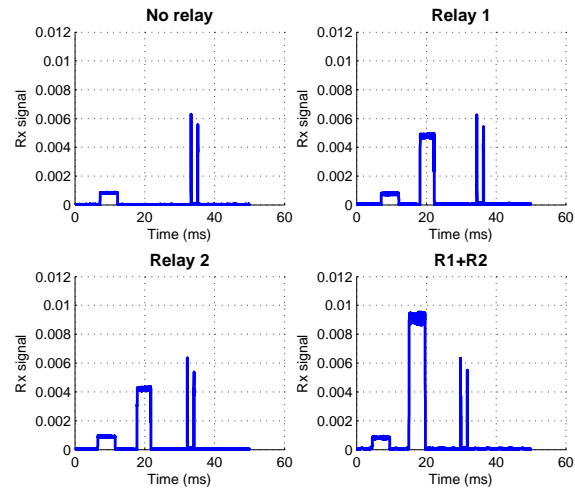}
	\caption{Received signal during one cycle of the setup with (a) no relays, (b) relay 1, (c) relay 2 and (d) relays 1 and 2 activated.}
	\label{fig:rx_ampl_1cycle}
\end{figure}

In this section we present results obtained with our experimental prototype. The prototype was run in an indoor environment, with a distance between transmitter and relay/receiver node of approximately 5~m. The nodes were kept static during the experiments, and there was little movement around the testbed to limit the effects of dynamic fading. The random phase perturbation are chosen randomly from the discrete set $\{-10^{\circ},10^{\circ}\}$.  
Figure \ref{fig:rx_ampl_1cycle} shows the received signal during a single cycle for two relay nodes. The relay delay time was set at $T_d=10$~ms, and the cycle period at $T_c = 50$~ms.  In subfigure~(a), no relay is activated: only the message from the (distant) transmitter is observed, with low amplitude. In subfigures~(b) and (c), only relay 1 or relay 2 is activated. After the original message from the transmitter, the (stronger) message from the relay can be observed. Finally, when both relays have been activated and convergence of the one-bit feedback algorithm has been achieved, it can be seen in subfigure~(d) that the relayed packets from both relays add up coherently. The amplitude of the relayed packets is then equal to the sum of the amplitudes of the individual relayed packets. Note that in all figures there is a noisy signal after the relayed packets. This corresponds to the self-interference created by the receiver's feedback message to the relays, in an adjacent frequency band, and can be ignored. 

\begin{figure}[ht]
	\centering
	\includegraphics[width=8.5cm]{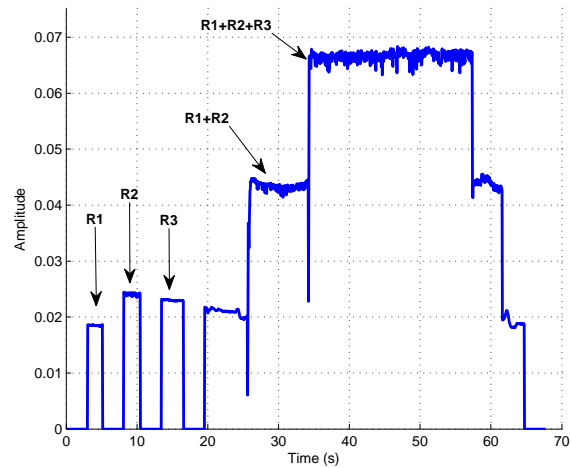}
	\caption{Mean amplitude of the relayed packets with 3 relay nodes. }
	\label{fig:rxAmpl_3relays}
\end{figure}

\begin{figure}[ht]
	\centering
	\includegraphics[width=8.5cm]{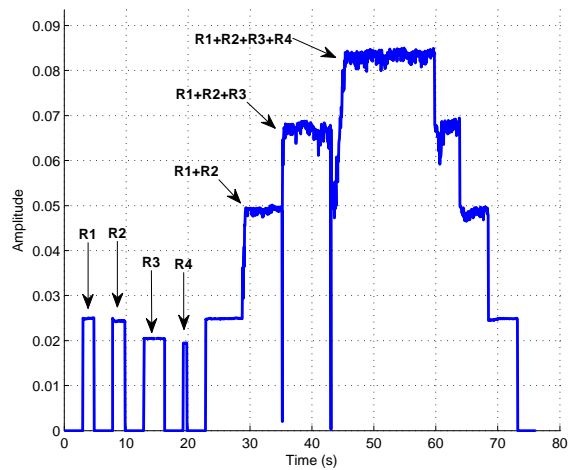}
	\caption{Mean amplitude of the relayed packets with 4 relay nodes. }
	\label{fig:rxAmpl_4relays}
\end{figure}

Figures \ref{fig:rxAmpl_3relays} and \ref{fig:rxAmpl_4relays} show the mean amplitude of the relayed packets only, over longer amounts of time, using 3 relays and 4 relays, respectively. It can be seen that the amplitude of the combined relayed messages correspond to the sum of the amplitudes of the individual relayed messages. Also, it can be observed that, once the one-bit feedback achieves convergence, the amplitude of the relayed messages is stable at its maximum value.
Thus, the phase errors due to LO drift are being successfully handled by the one-bit feedback algorithm. 
In Figure \ref{fig:rxAmpl_4relays}, the steady increase in RSS can be observed when the 4th relay is turned on. A few iterations were necessary for the RSS to converge to its maximum value. It can also be seen in Figures \ref{fig:rxAmpl_3relays}-\ref{fig:rxAmpl_4relays} that there are slight dips once the RSS has converged to its maximum value. This is because the one-bit feedback algorithm continues running even after the RSS has achieved its maximum value, causing the phases to misalign and realign over time. An easy improvement would be to reduce the size of the random phase perturbation applied at the relays once the RSS converges to its maximum value. 

\medskip

\begin{figure}[ht]
	\centering
	\includegraphics[width=8.5cm]{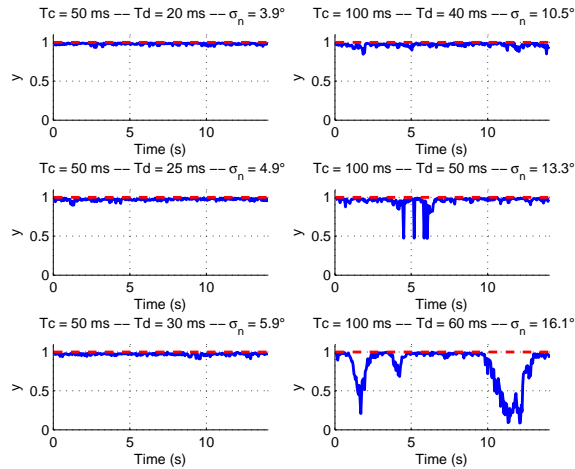}
	\caption{Mean amplitude of the relayed packets when varying $T_d$ and $T_c$. }
	\label{fig:varyingTcAndTd}
\end{figure}

In Section \ref{sec:phase_error} it was determined that increasing the relay delay time $T_d$ and the cycle period $T_c$ would result in an increasing phase error. In addition, it was argued in Section \ref{sec:one_bit_fb} that if the phase error becomes large (with respect to the size of the random phase perturbation), the RSS drift will decrease and the one-bit feedback algorithm will not be able to maintain the amplitude at its theoretical maximum. To verify these predictions, our experimental testbed was run with different values of $T_c$ and $T_d$, as shown in Figure~\ref{fig:varyingTcAndTd}. The setup was run with two relay nodes, and the random phase perturbation on both relay nodes was $10^{\circ}$. The LO parameters of our testbed were estimated previously \cite{Quitin:TWC:2013} as $q_1^2=8.47\times 10^{-22}$ and $q_2^2=5.51\times 10^{-18}$. The corresponding phase error standard deviation, computed using \eqref{eq:inter_cycle_var}, is given in the title of the subfigures. For each test, we first waited for a period of time long enough that the one-bit feedback algorithm could be expected to converge. The red line represents the (normalized) maximum possible RSS (based on the measured amplitudes of the relayed packets when the relays are turned on individually), and the blue line corresponds to the (normalized) measured RSS of the relayed packets when both relays are turned on, after convergence of the one-bit feedback algorithm. It can be seen that once the phase error standard deviation becomes significant with respect to that of the random phase perturbation, the one-bit feedback algorithm has trouble converging, and the RSS has trouble maintaining its maximum value. 
\section{Conclusions}
\label{sec:conclusion}

Starting from the observation that over-the-air combining using amplify-forward relaying provides a scalable approach to distributed receive beamforming,
we have proposed an architecture for achieving the synchronization required for the relayed signals to cohere at the receiver.  
An attractive feature of the time division (between long and short links) approach considered here is that frequency synchronization comes for free.
We have demonstrated this architecture using a software-defined radio testbed, and report experimental results achieving
the receive beamforming gains predicted by theory.  We also model and analyze the potential performance degradation due to phase errors accumulating
due to LO drift.  We provide an analytical framework, verified via Monte Carlo simulations, which estimates the degradation of the 
RSS attained by the one-bit feedback algorithm with finite memory in the presence of phase errors.  
A key insight, also verified experimentally, is that significant performance degradation occurs if the variance of the phase noise is comparable to, or larger than, the variance of the random phase perturbation used in the one-bit feedback algorithm.  This provides guidance on choice of system parameters
such as LO quality, relaying delay, and cycle length.  The open-source implementation of our prototype is publicly available, and hopefully provides a starting point for further implementation of solutions for distributed MIMO.

There are many directions for future work.  An important topic is generalization of our amplify-forward approach to provide
scalable distribution reception over wideband dispersive channels.
Possible approaches include ``filter-and-forward,'' or amplify-forward on a per-subcarrier basis.  Design challenges include
timing synchronization and tracking schemes, and the development of parsimonious feedback strategies.  Also, while our time division architecture
yields implicit frequency synchronization, there may be many scenarios in which frequency division between long and short links
is an attractive design choice, in which case explicit frequency synchronization is required.  Finally, it is important to develop and evaluate
designs that account for mobile nodes, possibly with different models addressing different potential applications.


\appendices

\section{Joint probability distribution of $U$ and $V$}
\label{sec:joint_prob_U_V}

\begin{figure}[ht]
	\centering
	\includegraphics[width=8.5cm]{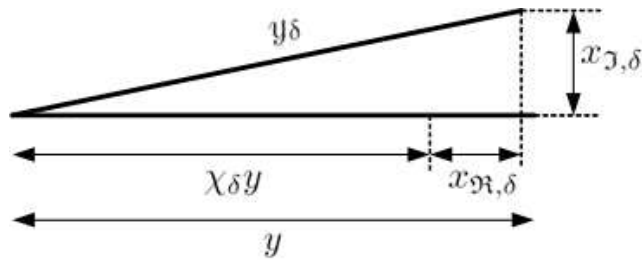}
	\caption{Effect of a random phase perturbation on the total received signal}
	\label{fig:effect_rand_ph_pert}
\end{figure}

It was shown in \cite{Mudumbai:TIT:2010} that for large $N$, the net effect of a random phase perturbation on the total signal can be modeled as shown in Figure~\ref{fig:effect_rand_ph_pert}. 
The effect of phase noise (or random phase perturbations plus phase noise) on the total signal can be modeled in an identical manner. Using equation (22) in \cite{Mudumbai:TIT:2010}, for large $N$ and $y$, the following approximation can then be made: 
\begin{align*}
	y_{\delta} &\approx \chi_{\delta} y + x_{\Re,\delta} \\
	y_{n} &\approx \chi_{n} y + x_{\Re,n} \\
	y_{\delta n} &\approx \chi_{\delta n} y + x_{\Re,\delta n} 
\end{align*}
The variables $x_{\Re,\delta}$, $x_{\Re,n}$ and $x_{\Re,\delta n}$ are zero-mean Gaussian random variables with variances $\sigma_{\Re,\delta}^2 = \frac{1-\chi_{\delta}^2-\rho_{\delta}\kappa(y)}{2N}$, $\sigma_{\Re,n}^2 = \frac{1-\chi_{n}^2-\rho_{n}\kappa(y)}{2N}$ and $\sigma_{\Re,\delta n}^2 = \frac{1-\chi_{\delta n}^2-\rho_{\delta n}\kappa(y)}{2N}$, respectively. It can be seen that the joint statistics of the random variables $U = y_{\delta n}-y_{n}$ and $V = y_{\delta}-y$ are simply those of a bivariate Gaussian distribution, entirely characterized by the means of $U$ and $V$, the variances of $U$ and $V$, and the covariance between $U$ and $V$. The development in the following subsections are similar to the development made in \cite{Mudumbai:TIT:2010}-Appendix~C. 

\subsection{Mean and variance of $U$}

From Figure~\ref{fig:effect_rand_ph_pert}, we can define the following terms
\begin{align*}
x_{\delta n} &= x_{\Re,\delta n}+j x_{\Im,\delta n} = \frac{1}{N}\sum\limits_{i=1}^{N} e^{j\phi_i}(e^{j(\delta_i+n_i')}-\chi_{\delta n}) \\
x_{n} 			 &= x_{\Re, n}+j x_{\Im, n} 						= \frac{1}{N}\sum\limits_{i=1}^{N} e^{j\phi_i}(e^{j n_i}-\chi_n) \\
x_{\delta}   &= x_{\Re,\delta}+j x_{\Im,\delta} 		= \frac{1}{N}\sum\limits_{i=1}^{N} e^{j\phi_i}(e^{j \delta_i}-\chi_{\delta}) 
\end{align*}
The mean and variance of $U = y_{\delta n}-y_{n}$ are then given by
\begin{align*}
\mathbb{E}[U] &= (\chi_{\delta n}-\chi_n)y \\
\text{Var}[U] &= \mathbb{E}\left[ (x_{\Re,\delta n}-x_{\Re, n})^2 \right] \\
				&= \frac{1}{4} \mathbb{E}\left[ \left( x_{\delta n}+x_{\delta n}^*-x_n-x_n^* \right)^2 \right] \\
 				&= \frac{1}{4} \mathbb{E}\left[ x_{\delta n}^2+x_{\delta n}^{*2}+x_n^2+x_n^{*2} +2x_{\delta n}x_{\delta n}^*+2x_nx_n^* -2x_{\delta n}x_n-2x_{\delta n}x_n^*-2x_{\delta n}^*x_n-2x_{\delta n}^*x_n^* \right] \\
\end{align*}
The first term of the previous equation is given by 
\begin{align*}
\mathbb{E}\left[x_{\delta n}^2\right] = \frac{1}{N^2}\sum\limits_{i=1}^{N}\sum\limits_{l=1}^{N} \mathbb{E}\left[ e^{j(\phi_i+\phi_l)} (e^{j(\delta_i+n_i')}-\chi_{\delta n}) (e^{j(\delta_l+n_l')}-\chi_{\delta n}) \right] 
\end{align*}
Since the phase perturbations $\delta_i$ and noise terms $n_i'$ have a symmetric distribution, it follows that the terms $(e^{j(\delta_i+n_i')}-\chi_{\delta n})$ and $(e^{j(\delta_l+n_l')}-\chi_{\delta n})$ have zero mean. Moreover, since the phase perturbations $\delta_i$ and noise terms $n_i'$ are independent for different nodes, the term in the sum are zero when $i\neq l$ and the previous becomes 
\begin{align*}
\mathbb{E}\left[x_{\delta n}^2\right] &= \frac{1}{N^2}\sum\limits_{i=1}^{N} \mathbb{E}\left[e^{j2\phi_i} (e^{j(\delta_i+n_i')}-\chi_{\delta n})^2 \right] \\
																			&= -\frac{1}{N^2} \rho_{\delta n} \sum\limits_{i=1}^{N} \mathbb{E}\left[e^{j2\phi_i} \right]
\end{align*}
where we used the approximation that, for small phase perturbations and/or phase noises, $\mathbb{E}\left[e^{j(\delta_i+n_i')}\right] \approx \mathbb{E}\left[\cos(\delta_i+n_i')\right]$. Similarly, one can compute 
\begin{align*}
\mathbb{E}\left[x_{\delta n}x_{\delta n}^* \right] 
		&= \frac{1}{N^2}\sum\limits_{i=1}^{N}\sum\limits_{l=1}^{N} \mathbb{E}\left[e^{j(\phi_i-\phi_l)} (e^{j(\delta_i+n_i')}-\chi_{\delta n}) (e^{-j(\delta_l+n_l')}-\chi_{\delta n}) \right] \\
		&= \frac{1}{N^2}\sum\limits_{i=1}^{N} \mathbb{E}\left[ \left| e^{j(\delta_i+n_i')} -\chi_{\delta n} \right|^2 \right] \\
		&= \frac{1}{N}(1-\chi_{\delta n}^2)
\end{align*}
since $\mathbb{E}\left[ \left| e^{j(\delta_i+n_i')} -\chi_{\delta n} \right|^2 \right]=1-\chi_{\delta n}^2$ (see \cite{Mudumbai:TIT:2010}). 
Following similar arguments, one can easily obtain the following expressions 
\begin{align*}
\mathbb{E}\left[x_{\delta n}^{*2}\right]  &= -\frac{1}{N^2} \rho_{\delta n} \sum\limits_{i=1}^{N} \mathbb{E}\left[e^{-j2\phi_i} \right] \\
\mathbb{E}\left[x_{n}^2\right]  			&= -\frac{1}{N^2} \rho_{n} \sum\limits_{i=1}^{N} \mathbb{E}\left[e^{j2\phi_i} \right] \\
\mathbb{E}\left[x_{n}^{*2}\right]  			&= -\frac{1}{N^2} \rho_{n} \sum\limits_{i=1}^{N} \mathbb{E}\left[e^{-j2\phi_i} \right] \\
\mathbb{E}\left[x_{n}x_{n}^* \right]  &= \frac{1}{N}(1-\chi_{n}^2)
\end{align*}
The crossed terms are computed as
\begin{align*}
\mathbb{E}\left[x_{\delta n}x_{n} \right] 
		&= \frac{1}{N^2}\sum\limits_{i=1}^{N}\sum\limits_{l=1}^{N} \mathbb{E}\left[e^{j(\phi_i+\phi_l)}(e^{j(\delta_i+n_i')}-\chi_{\delta n}) (e^{j(n_l)}-\chi_{n})\right]\\
\end{align*}
The terms $(e^{j(\delta_i+n_i')}-\chi_{\delta n})$ and $(e^{j(n_l)}-\chi_{n})$ are both zero-mean and independent variables, and therefore all terms in the summation equal zero, which leads to
\begin{align*}
\mathbb{E}\left[x_{\delta n}x_{n} \right] = \mathbb{E}\left[x_{\delta n}x_{n}^* \right] = \mathbb{E}\left[x_{\delta n}^*x_{n} \right] = \mathbb{E}\left[x_{\delta n}^*x_{n}^* \right] = 0
\end{align*}
The terms $\kappa(y)=\frac{1}{N}\sum\limits_{i=1}^{N}\mathbb{E}[e^{2j\phi_i}]$ depends on $y$ only, and can be approximated by $\kappa(y)=e^{-4(1-y)}$ for large $y$. The latter was derived in Conjecture 1 in \cite{Mudumbai:TIT:2010}), where statistical mechanics arguments where used to determine that that the distribution of the phases around their mean follow an Exp-Cosine distribution. 

\medskip

The variance of $U$ can finally be written as 
\begin{align*}
\text{Var}[U] &= \frac{1-\chi_{\delta n}^2-\rho_{\delta n}\kappa(y)}{2N} + \frac{1-\chi_{n}^2-\rho_{n}\kappa(y)}{2N}
\end{align*}

\subsection{Mean and variance of $V$}

The statistics of $V = y_{\delta}-y$ can be deduced in a manner identical to the statistics of $U$, which leads to the following expressions: 
\begin{align*}
\mathbb{E}[V] &= (\chi_{\delta}-1)y \\
\text{Var}[V] &= \frac{1-\chi_{\delta}^2-\rho_{\delta}\kappa(y)}{2N} 
\end{align*}

\subsection{Covariance between $U$ and $V$}

The two terms $U$ and $V$ are not independent, but the covariance between both can easily be obtained following similar arguments as before. 
\begin{align*}
\text{Cov}[U,V] &= \mathbb{E}[(x_{\Re,\delta n} - x_{\Re,n})x_{\Re,\delta}] \\
 								&= \frac{1}{4}\mathbb{E}\left[ \left(x_{\delta n}+x_{\delta n}^*-x_n-x_n^*\right) \left(x_{\delta}+x_{\delta}^*\right)  \right] \\
 								&= \frac{1}{4}\mathbb{E}\left[ x_{\delta n}x_{\delta} + x_{\delta n}x_{\delta}^* + x_{\delta n}^*x_{\delta} + x_{\delta n}^*x_{\delta}^* -x_nx_{\delta} -x_nx_{\delta}^* - x_n^*x_{\delta} - x_n^*x_{\delta}^*   \right]
\end{align*}
Note that the mean of the last four terms can immediately seen to be zero, since the phase perturbations terms and the noise terms are zero-mean and independent. The first term can be computed as follows: 
\begin{align*}
\mathbb{E}\left[ x_{\delta n}x_{\delta} \right]
	&= \frac{1}{N^2}\sum\limits_{i=1}^{N}\sum\limits_{l=1}^{N} \mathbb{E}\left[ e^{j\phi_i}(e^{j(\delta_i+n_i')}-\chi_{\delta n}) e^{j\phi_l}(e^{j\delta_l}-\chi_{\delta})\right] \\
	&= \frac{1}{N^2}\sum\limits_{i=1}^{N}\mathbb{E}\left[e^{2j\phi_i} \left(e^{j(2\delta_i+n_i)}-\chi_{\delta n}e^{j\delta_i}-\chi_{\delta}e^{j(\delta_i+n_i)}+\chi_{\delta n}\chi_{\delta}\right) \right] \\
  &= \frac{1}{N^2}\sum\limits_{i=1}^{N}\mathbb{E}\left[ e^{2j\phi_i} \right] \left( \mathbb{E}[\cos(2\delta_i+n_i)] - \chi_{\delta n}\chi_{\delta} - \chi_{\delta}\chi_{\delta n}+\chi_{\delta n}\chi_{\delta}    \right) \\
  &= \frac{1}{N^2}\sum\limits_{i=1}^{N}\mathbb{E}\left[ e^{2j\phi_i} \right] \left( \mathbb{E}[\cos(2\delta_i+n_i)] - \chi_{\delta n}\chi_{\delta}\right)
\end{align*}
Similarly, one can compute that
\begin{align*}
\mathbb{E}\left[ x_{\delta n}^*x_{\delta}^* \right]
  &= \frac{1}{N^2}\sum\limits_{i=1}^{N}\mathbb{E}\left[ e^{-2j\phi_i} \right] \left( \mathbb{E}[\cos(2\delta_i+n_i)] - \chi_{\delta n}\chi_{\delta}  \right)
\end{align*}
The crossed terms can be computed as follows
\begin{align*}
\mathbb{E}\left[ x_{\delta n}x_{\delta}^* \right]
	&= \frac{1}{N^2}\sum\limits_{i=1}^{N}\sum\limits_{l=1}^{N} \mathbb{E}\left[ e^{j\phi_i}(e^{j(\delta_i+n_i')}-\chi_{\delta n}) e^{-j\phi_l}(e^{-j\delta_l}-\chi_{\delta})\right] \\
	&= \frac{1}{N^2}\sum\limits_{i=1}^{N}\mathbb{E}\left[ (e^{j(\delta_i+n_i')}-\chi_{\delta n})(e^{-j\delta_i}-\chi_{\delta}) \right] \\
	&= \frac{1}{N^2}\sum\limits_{i=1}^{N}\mathbb{E}\left[ e^{jn_i'} - \chi_{\delta}e^{j(\delta_i+n_i')} - \chi_{\delta n}e^{-j\delta_i} + \chi_{\delta n}\chi_{\delta}  \right] \\
	&= \frac{1}{N}\left(\chi_n - \chi_{\delta n}\chi_{\delta}\right)
\end{align*}
and 
\begin{align*}
\mathbb{E}\left[ x_{\delta n}^*x_{\delta} \right]
	&= \frac{1}{N}\left(\chi_n - \chi_{\delta n}\chi_{\delta}\right)
\end{align*}

\medskip

Finally, the covariance between $U$ and $V$ can be written as
\begin{align*}
\text{Cov}[U,V] &= \frac{\chi_n-\chi_{\delta n}\chi_{\delta} - \rho_{2\delta n}\kappa(y)}{2N}
\end{align*}
where $\rho_{2\delta n} = \chi_{\delta n}\chi_{\delta} - \mathbb{E}[\cos(2\delta_i+n_i)]$.

\section*{Acknowledgements}
This work is funded in part by the US National Science Foundation under grant CCF-1302114, and by the Institute for Collaborative Biotechnologies through grant W911NF-09-0001 from the U.S. Army Research Office. The content of the information does not necessarily reflect the position or the policy of the Government, and no official endorsement should be inferred.

\footnotesize{
\bibliographystyle{IEEEtran}
\bibliography{references}
}

\end{document}